\documentclass{article}
\usepackage{locata,amsmath,graphicx,url,times, subcaption}

\title{Binaural Source Localization based on Modulation-Domain Features and Decision Pooling}
\name{Semih A\u{g}caer and Rainer Martin }
\address{Ruhr-Universit\"{a}t Bochum,    Institute of Communication Acoustics, Bochum, Germany\\
	\{semih.agcaer, rainer.martin\}@rub.de
}
\pdfoutput=1 
\begin{document}

\maketitle


\begin{abstract}
	In this work we apply Amplitude Modulation Spectrum (AMS) features to the source localization problem. Our  approach computes 36 bilateral features for 2s long signal segments and estimates the azimuthal directions of a sound source through a binaurally trained classifier. This directional information of a  sound source could be e.g. used to steer the beamformer in a hearing aid to the source of interest in order to increase the SNR. We evaluated our approach on the development set of the  IEEE-AASP Challenge on sound source localization and  tracking (LOCATA) and achieved a  4.25$^\circ$ smaller MAE than the baseline approach. Additionally, our approach is  computationally less complex.    
\end{abstract}


\section{Introduction}
\label{sec:intro}
Source localization is an important task in many acoustic signal processing applications. An estimated azimuthal source direction could be e.g. incorporated in an acoustic signal processing system using an adaptive beamformer to steer the beamformer into the direction of  the source and thus to improve the SNR. \\
In our previous work, we proposed an Amplitude Modulation Spectrum (AMS) based feature extraction algorithm which we successfully used for acoustic scene classification \cite{AM15}. The feature extraction is based mainly on two successive filter banks, non-linear operations, and a final averaging step. In this paper, we apply these amplitude modulation features to the speaker localization task. For this we extract 4x9 features from the four cardioid signals of hearing aid microphones. With these 36 features a source azimuth is estimated for each 2s long signal segment by pooling decisions from a set of six classifiers.\\ 
In contrast to other common approaches, we do not compute any interaural phase/time difference  (IPD/ITD) and/or interaural level differences (ILD) explicitly like in \cite{Puder12, Ying13, Lee2010, MZ2018}. We compute nine AMS-based features separately for each cardioid signal and then concatenate them to a feature vector with a length of 36.  However, the 36 AMS features will implicitly contain ILD information. It should be noted that AMS-based features are inspired by the human auditory system and are successfully used for acoustic scene classification tasks \cite{AM15, AMS2008, AMS2010, Langner, AMS2011}.\\
We evaluate the performance of the proposed method on the development set of the LOCATA Challenge \cite{LOCATA2018a}. The LOCATA Challenge offers six different tasks from which we chose the first one, which is localization of a single static source with a static microphone array.\\
In the next sections of this paper, we will recap  our AMS-based features, describe our classification system and our  training data. Then, in Section 3, we  evaluate our proposed approach on the LOCATA Challenge corpus and present the  results. Section 4 concludes this paper with a discussion of these results.

\begin{figure}[t]
	\centering
	\resizebox{0.4\textwidth}{!}
	{\includegraphics[width = 0.9\textwidth]{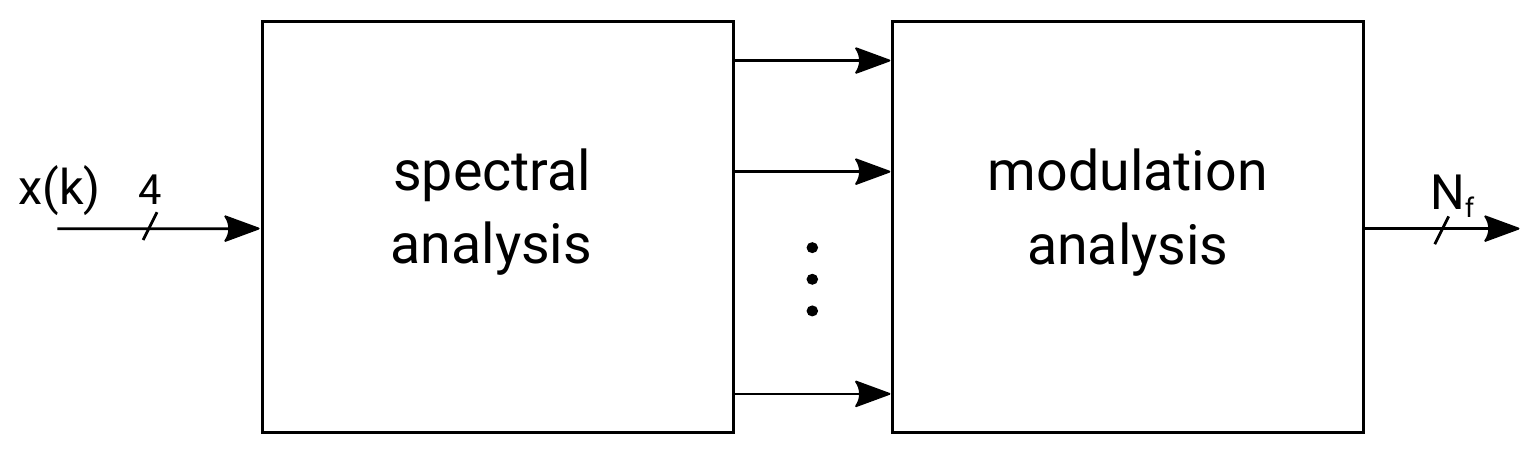}} 
	\caption{AMS based modulation-domain feature extraction system}
	\label{Fig_AMS_block}
\end{figure}

\section{Methods}
\label{sec:methods}

\subsection{Modulation-Domain Feature Extraction}
Our  feature extraction approach is similar to the approach we proposed for the DCASE Challenge 2013 \cite{AASPDataSet} which is described in detail in \cite{AM15}.  The  feature extraction is depicted in Figure \ref{Fig_AMS_block} and it mainly consist of two blocks. First, the input signals are spectrally analyzed by $N_s$ bandpass filters  and then the amplitude modulation in each spectral bins are analyzed by $N_m$ different filters and averaged over the frame length $T_s$. \\
For the LOCATA Challenge we set   $N_s = 3$ and $N_m = 3$ to three  and the frame length $T_s = 2$s. Instead of using a single microphone signal for the DCASE challenge we use the microphone signals of hearing aid dummies and extract four cardioid signals (front left, front right, back left and back right) by feeding the  microphone signals into the respective beamformers. These four cardiod signals are then fed in parallel into the AMS feature extraction algorithm which outputs   $N_f =4 \cdot N_s \cdot N_m = 36$ features for one frame. \\
The  passband range of each filter in the filterbanks is found by an iterative optimization method. We optimize the passband ranges by a model-based optimization (MBO) approach. The MBO is an iterative approach for the optimization of a black box objective function. It is used when the evaluation of an objective function, in our case the classification error depending on different filter bank parameters, is expensive in terms of  available resources (computational time). MBO tries to construct an approximation model, a so called surrogate model, of this expensive objective function to find the optimal parameter for a given problem. The evaluation of the surrogate model is cheaper than the original objective function.  For a more detailed description of MBO we refer to \cite{Forrester, Weihs}.\\

\subsection{Classification and Decision Pooling} 
\begin{figure}[t]
	\centering
	\resizebox{0.5\textwidth}{!}
	{\includegraphics[width = 0.9\textwidth]{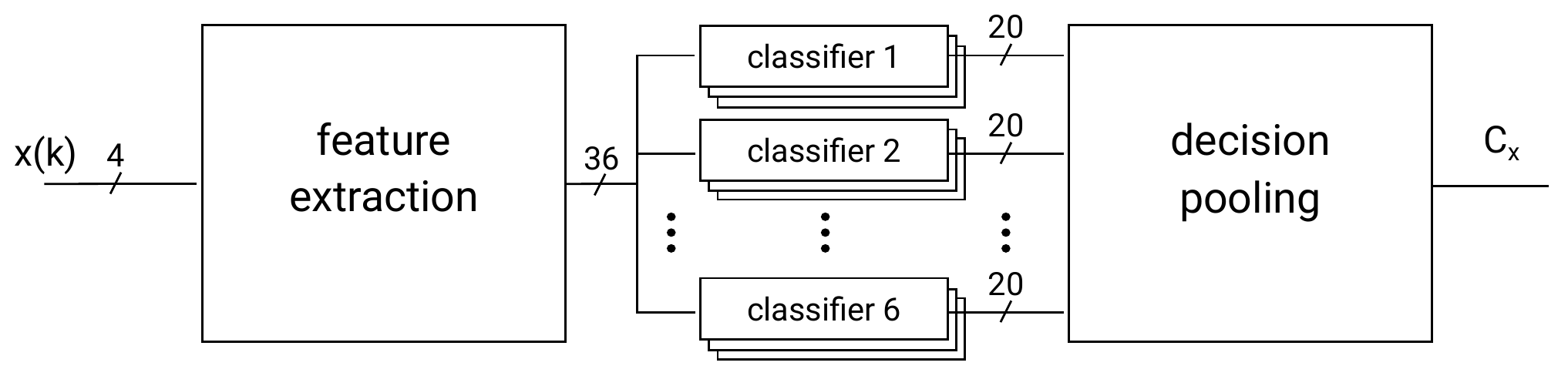}} 
	\caption{Classification system based on decision pooling.}
	\label{Fig_classifier}
\end{figure}

\begin{figure}[t]
	\centering
	\begin{subfigure}[c]{4.2cm}
		\includegraphics[width = 0.9\textwidth]{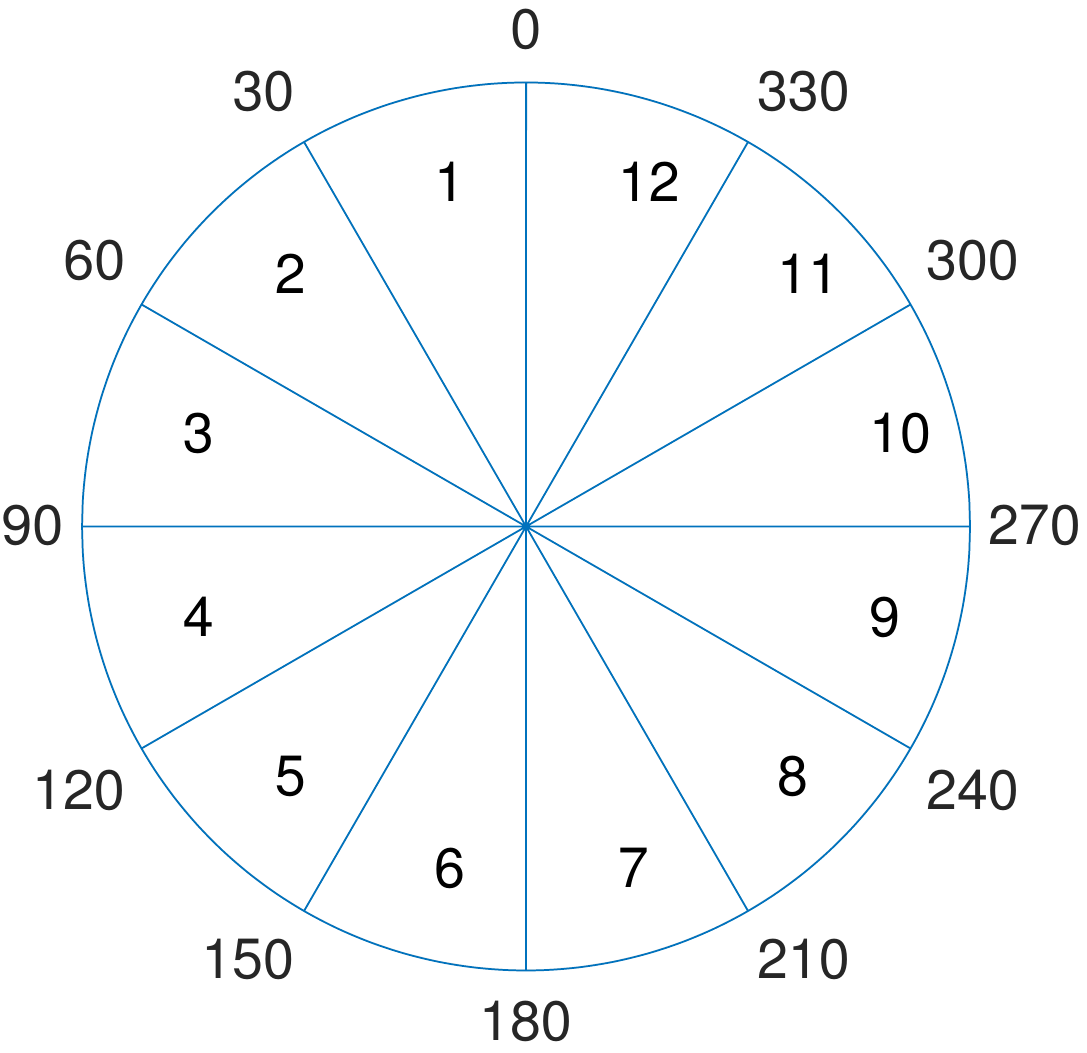}
		\subcaption{classifier set 1}
	\end{subfigure}
	\begin{subfigure}[c]{4.2cm}
		\includegraphics[width = 0.9\textwidth]{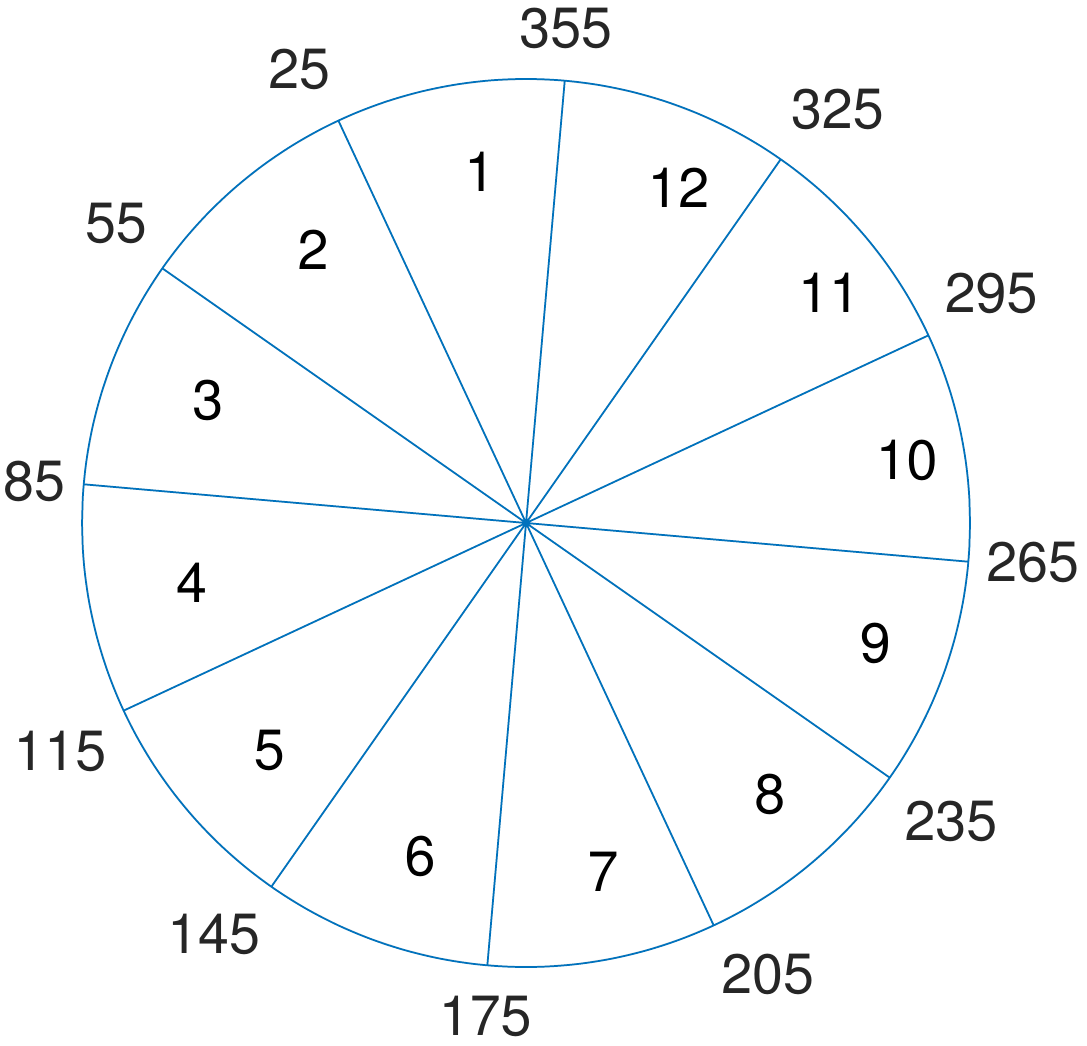}
		\subcaption{classifier set 2}
	\end{subfigure}
	\begin{subfigure}[c]{4.2cm}
		{\includegraphics[width = 0.97\textwidth]{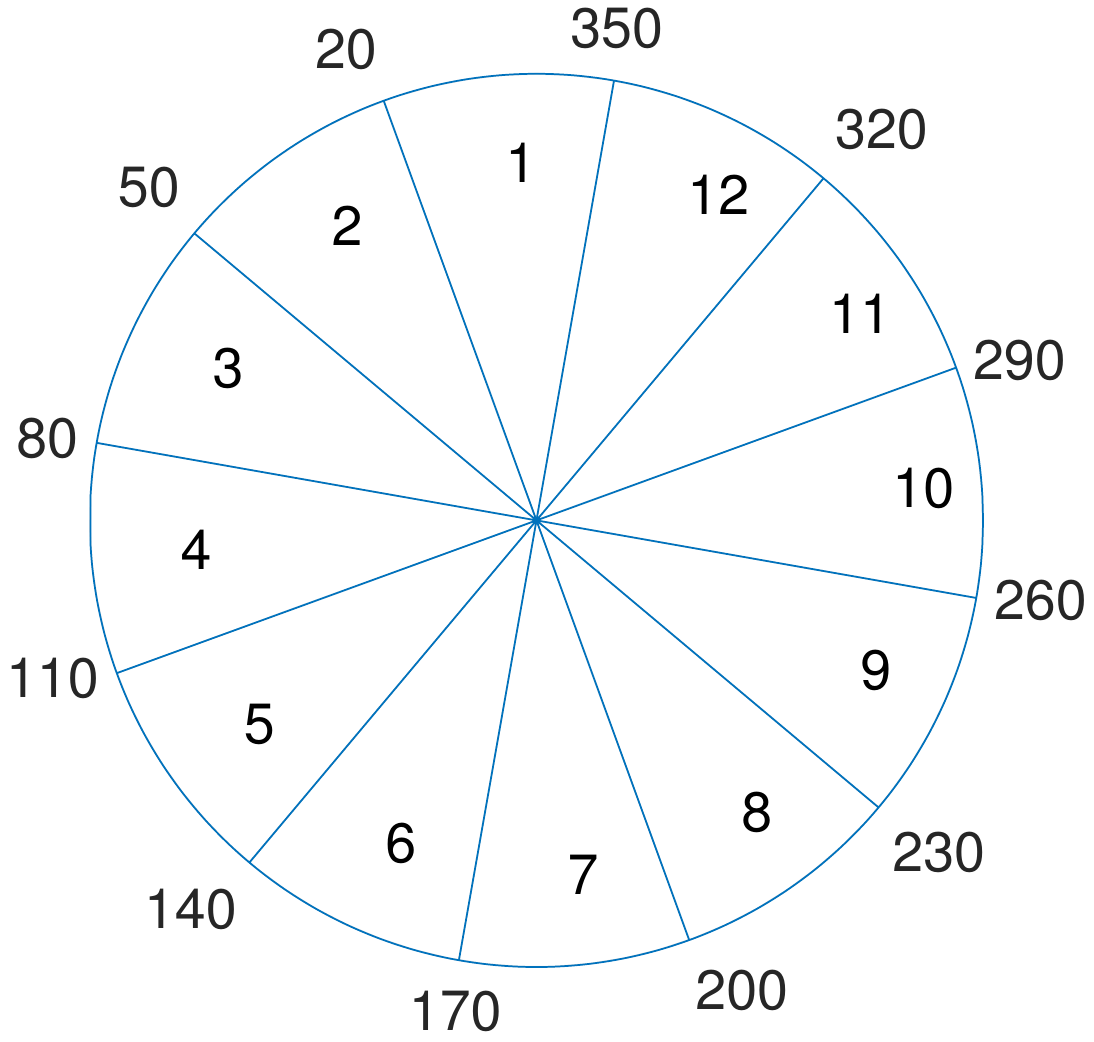}} 
		\subcaption{classifier set 3}
	\end{subfigure}
	\begin{subfigure}[c]{4.2cm}
		{\includegraphics[width = 0.97\textwidth]{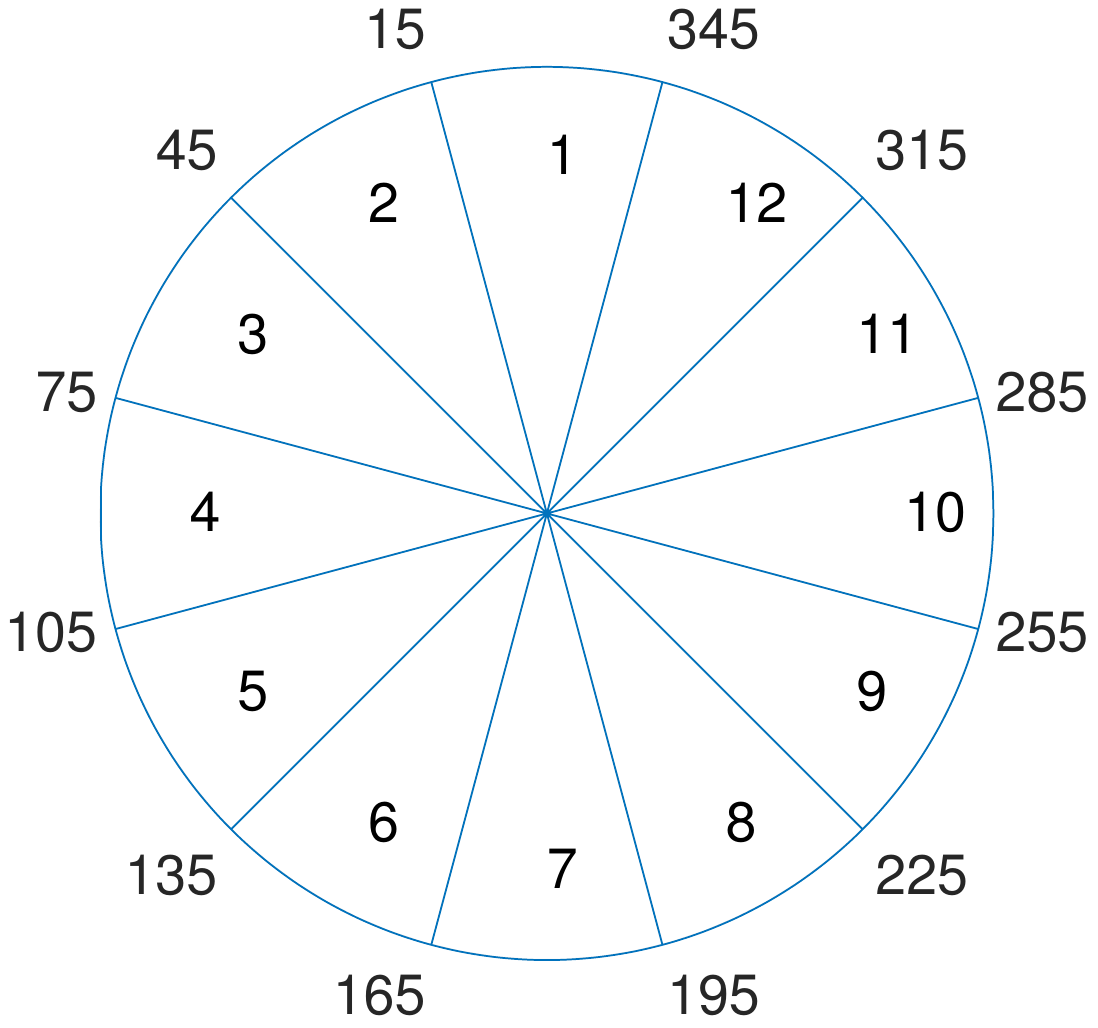}} 
		\subcaption{classifier 4}
	\end{subfigure}
	\begin{subfigure}[c]{4.2cm}
		{\includegraphics[width = 0.97\textwidth]{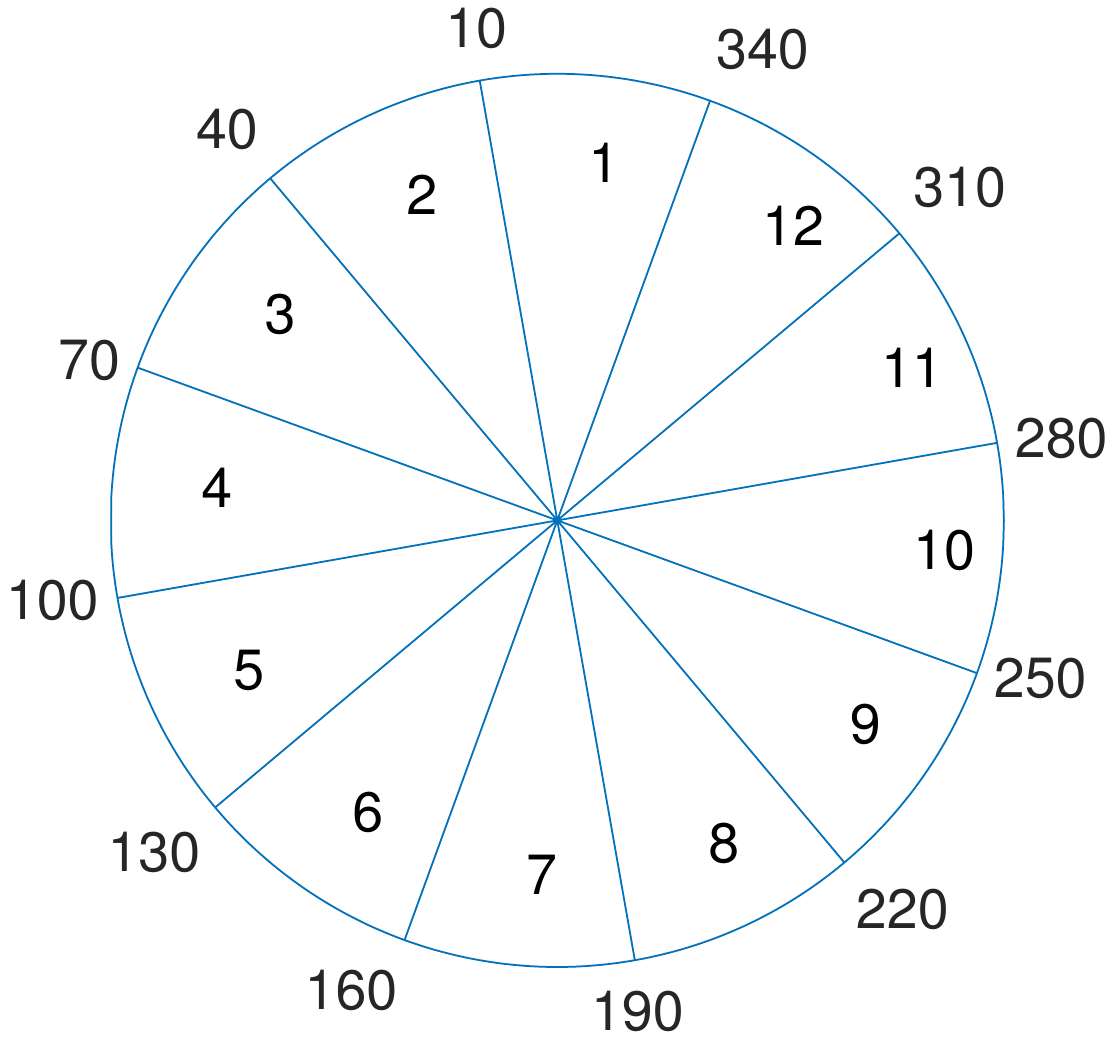}} 
		\subcaption{classifier set 5}
	\end{subfigure}
	\begin{subfigure}[c]{4.2cm}
		{\includegraphics[width = 0.97\textwidth]{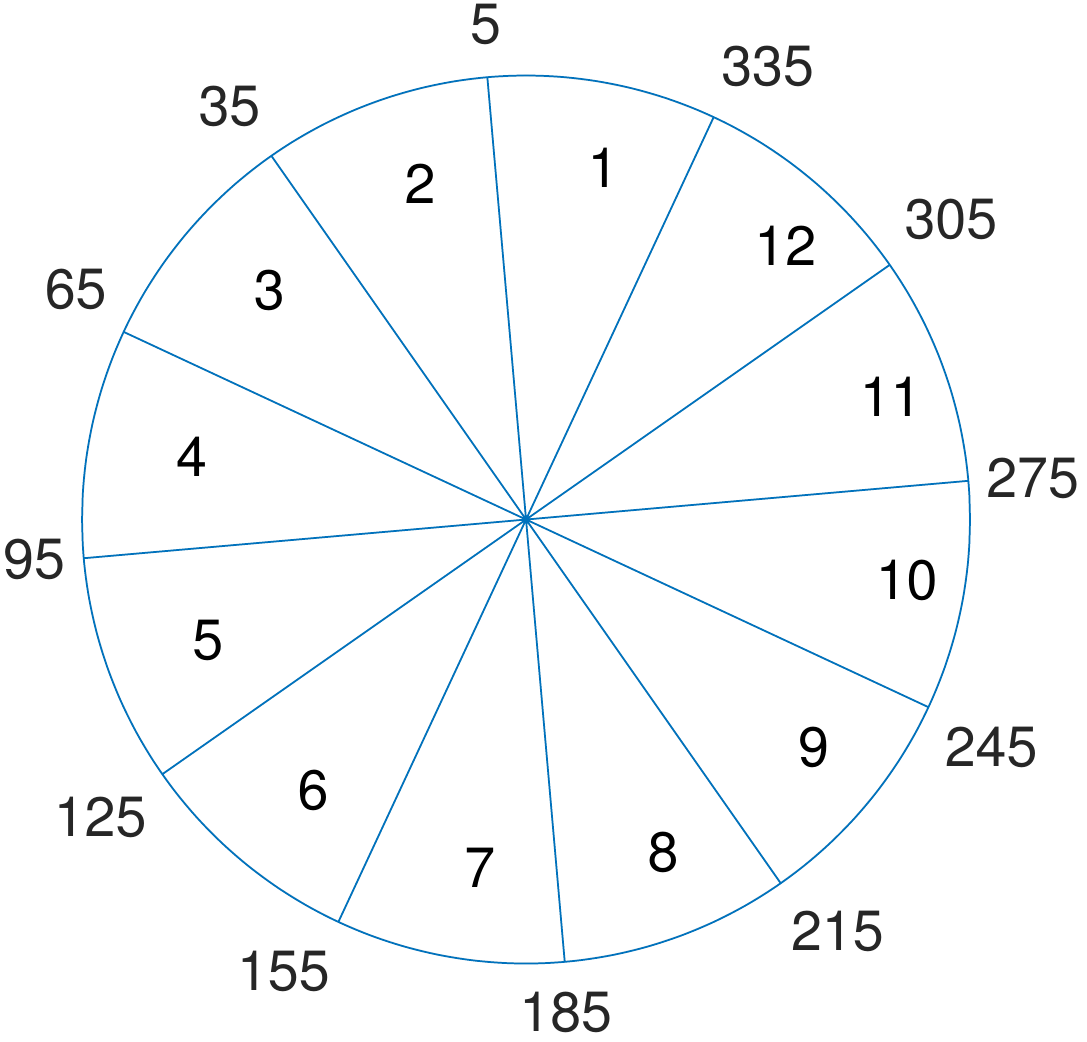}} 
		\subcaption{classifier set 6}
	\end{subfigure} 
	\caption{Class definition of the 6 classifier sets in degrees. Each classifier can distinguish between12 classes, where each class covers a range of 30$^\circ$}.
	\label{Fig_ClassifierDef}
\end{figure}

Our classification system, which is depicted in Figure \ref{Fig_classifier}, consist of six classifier sets  each with a resolution of 30$^\circ$. Any individual classifier can therefore distinguish between 12 different classes. The class definitions for each set is different and is depicted in Figure \ref{Fig_ClassifierDef}.  The six class definitions differ by 5$^\circ$ shifts leading to 72 discrete speaker directions and a theoretical final speaker localization resolution of  5$^\circ$ if all six classifiers results are combined. The classifier used in our system is a Linear Discriminant Analysis (LDA) classifier.\\
For each 2s signal frame we run five  4-fold cross-validations for each classifier set leading to $20 \cdot 6 = 120$ classifiers in total. In all classifiers we fed in the same 36 AMS features we extracted before.  The final speaker localization for one recording was determined by pooling the classification decision for each of the 72 azimuth bins over all 120 classification results for each frame in one recording and average the two most frequent azimuth estimates. 


\subsection{Training Data}
For training our classification system we created a data set with a speaker from 72 directions spaced in 5$^\circ$ steps i.e. from 0$^\circ$, 5$^\circ$, \dots, 355$^\circ$. The data set was rendered  using the Oldenburg HRTF data set (behind-the-ear hearing aids)\cite{Oldenburg} and various speech and noise data sets. For each direction we created 18 x 10s audio files  with variation in SNR level, speaker and noise type, which led to training set with a total duration of 3h 36min.

\section{Results}
\label{sec:results}
The development data set of the LOCATA Challenge consists of three recordings for the hearing aid microphone configuration. The audio signals have a sampling frequency of 48kHz which were  decimated to 20kHz. The left front, left back, right front and right back microphones of the hearing aid  were fed into a beamformer leading to four cardioid signals. From this four cardioid signals we compute our 36 features. We run our experiments on a PC with Intel(R) Core(TM) i5-3470 CPU @ 3.20GHz with 12GB RAM and Matlab 2017b 64-Bit. 
\begin{table}
	\centering
	\begin{tabular}{|c|c|}
		\hline 
		Recording	&Azimuth error in degree 				  \\  \hline 
		4	 & -9.86$^\circ$ 	   \\  \hline 
		5	 & -2.61$^\circ$ 	  \\  \hline 
		6	 & +2.39$^\circ$      \\  \hline 
		\hline
		MAE	 & 4.95$^\circ$\\  \hline      
	\end{tabular} 
	\caption{Azimuth error for each recording and the mean absolute error (MAE) for the development set. }
	\label{tab_results}
\end{table}
Table \ref{tab_results} shows the azimuth error for each recording in the development data set. We compute a single estimate for  each whole audio file by pooling over 2s long frames and assigned these to the required time stamps of the audio database. The achieved mean absolute error (MAE)  is  4.95$^\circ$, which is  4.25$^\circ$ better than the baseline MUSIC algorithm.\\
In order to compare the computational efficiency of our approach to the baseline MUSIC algorithm we  used the real time factor (RTF) as a measure, which is defined as the computation time for one recording divided by the duration of the recording. The mean RTF for the baseline MUSIC algorithm is 9.1 on our hardware setup, which means that it is not real time capable. In contrast, our proposed speaker localization approach has a mean RTF  of 0.7 and is thus real time capable. It is 13 times faster than the baseline MUSIC algorithm.

\section{Discussion and Conclusion}
\label{sec:discussion}
In this paper, we showed that our AMS feature extraction algorithm, which was used for audio scene classification before \cite{AM15}, could be also adapted for speaker localization. We trained our system with a  training set rendered with the Oldenburg HRTF set and evaluated our trained system on the publicly available LOCATA challenge development data set with one static speaker. The mean absolute azimuth estimation error of 4.95$^\circ$ is  by 4.25$^\circ$ lower than the baseline with the additional advantage of being  computationally less complex. It is up to 13 times faster and with a RTF of 0.7 real time capable on PC hardware. \\
Our system does not explicitly computed any IPD, ITD or ILD  informations. This makes it less dependent on the head-size and also eliminates the problem of phase synchronization which is required for computing  IPD/ITD.  However, we assume that our features implicitly contain ILD information in the modulation patterns.
\bibliographystyle{IEEEtran}
\bibliography{refs}
%
%
%
%
%
%
%
%
%

\end{document}